\documentclass[aps,floatfix,twocolumn,showpacs,superscriptaddress,groupedaddress]{revtex4}

\usepackage{amsmath,latexsym}
\usepackage{float}

\usepackage{graphicx}
\usepackage{epsfig}
\usepackage{color}

\def\Vf{V_{\mathrm{f}}}
\def\Vo{V_0}
\def\Piext{\Pi_\mathrm{ext}}
\def\Wdsc{W_{\mathrm{DSC}}}
\def\Wosm{W_\mathrm{osm}}

\def\dUVo{\Delta U_{V_0 \rightarrow V_{\mathrm{f}}}}  
\def\dUV_{\Delta U_{V \rightarrow V_\mathrm{f}}}  
\def\dUT0{\Delta U_{T_0}}  

\begin{document}

\title{Mechanics from Calorimetry: Probing the elasticity Elasticity for Responsive Hydrogels}

\author{Frank J. Aangenendt} 
\affiliation{Dutch Polymer Institute (DPI), P.O. Box 902, 5600AX Eindhoven, The Netherlands}
\affiliation{Institute for Complex Molecular Systems, Eindhoven University of Technology, P.O. Box 513, 5600MB Eindhoven, The Netherlands}  
\affiliation{Department of Mechanical Engineering, Materials Technology, Eindhoven University of Technology, P.O. Box 513, 5600MB Eindhoven, The Netherlands}  
\author{Johan Mattsson}  
\affiliation{School of Physics and Astronomy, University of Leeds, Leeds LS2 9JT, U.K.}  
\author{Wouter G. Ellenbroek} 
\affiliation{Institute for Complex Molecular Systems, Eindhoven University of Technology, P.O. Box 513, 5600MB Eindhoven, The Netherlands}  
\affiliation{Department of Physics, Eindhoven University of Technology, P.O. Box 513, 5600MB Eindhoven, The Netherlands}  
\author{Hans M. Wyss} 
\email{H.M.Wyss@tue.nl}  
\affiliation{Dutch Polymer Institute (DPI), P.O. Box 902, 5600AX Eindhoven, The Netherlands}  
\affiliation{Institute for Complex Molecular Systems, Eindhoven University of Technology, P.O. Box 513, 5600MB Eindhoven, The Netherlands}  
\affiliation{Department of Mechanical Engineering, Materials Technology, Eindhoven University of Technology, P.O. Box 513, 5600MB Eindhoven, The Netherlands}  
\begin{abstract}
\noindent  
Temperature-sensitive hydrogels based on polymers such as poly(\textit{N}-isopropylacrylamide) (PNIPAM) undergo a volume phase transition in response to changes in temperature. 
During this transition, distinct changes in both thermal and mechanical properties are observed. 
Here we illustrate and exploit the inherent thermodynamic link between thermal and mechanical properties by showing that the compressive elastic modulus of PNIPAM hydrogels can be probed using differential scanning calorimetry.    
We validate our approach by using conventional osmotic compression tests. 
Our method could be particularly valuable for determining the mechanical response of thermosensitive submicron-sized and/or oddly shaped particles, to which standard methods are not readily applicable.
\end{abstract}
\date{\today} 

\maketitle 
 
\begin{center}\rule{3in}{0.4pt}\end{center}

\section{Introduction}
\label{introduction}
Stimuli-responsive hydrogels respond to changes in their physical and chemical environment. 
The properties of these materials, including their volume and their mechanical response, can be reversibly controlled by external stimuli such as \textit{p}H, ionic concentration, temperature or electric fields~\citep{Wu:2003ue,Saunders:1996vt,Arleth:2005,Yamazaki1999103,Kim1999}.
Typical examples of this class of materials are hydrogels or microgels made from the polymer poly(\textit{N}-isopropylacrylamide) (PNIPAM), which exhibits a lower critical solution temperature (LCST) at around 32$^\circ\mathrm{C}$.
Near the LCST, a temperature variation of merely a few degrees can lead to a change in volume by more than an order of magnitude~\citep{Wu:2003ue,Saunders:1996vt,Arleth:2005}.
Such a dramatic response to external stimuli can be exploited in applications including stimuli-responsive surfaces~\citep{Liu201012377}, soft valves~\citep{Zhu20124017}, and responsive materials for drug delivery systems~\citep{Yamazaki1999103}.
The change in volume is only one manifestation of the alterations in the physical state of the system as the temperature is increased across the LCST; the thermal and mechanical properties of the material also undergo significant changes.

The associated changes in \emph{thermal properties} have been widely investigated in PNIPAM solutions and networks using differential scanning calorimetry (DSC)~\citep{APP:APP41669,Afroze200055}. 
Typically, a pronounced endothermic peak corresponding to the LCST is observed in the heat flow.
For PNIPAM microgels, alterations of the \emph{mechanical properties} have been investigated using multiple techniques, including conventional mechanical compression tests~\citep{Yuan2015164}, osmotic compression measurements~\citep{PhysRevE.84.011406,Sierra-Martin2011}, atomic force microscopy ~\citep{Hashmi:2009bv}, and Capillary Micromechanics~\citep{Wyss:2010iu,Guo:2011di,Voudouris:2013bc}.
For materials that can be considered elastically homogeneous, Capillary Micromechanics provides experimental access to both the elastic shear modulus $G$ and the compressive elastic modulus $K$, thus quantifying the full elastic behavior of the material, including the Poisson's ratio $\nu$.
Using this technique on PNIPAM microgels, a pronounced dip in the Poisson's ratio was observed around the LCST of these microgels~\citep{Voudouris:2013bc}, in agreement with previous measurements on macroscopic PNIPAM hydrogels by Hirotsu~\citep{Hirotsu:1990uj}, where an even more dramatic dip was observed.
This dip is due to a reduction of the compressive modulus $K$ relative to the shear modulus $G$, which can be rationalized by considering that the large thermal expansion coefficient near the LCST should intuitively go hand in hand with a large compressibility; the argument being that volumetric changes, whether thermally or mechanically induced, are less energetically costly in this temperature range.
This example illustrates the inherent connection between the thermal and mechanical behavior of these materials. 

Indeed, knowledge of the mechanical properties of responsive materials is often of key importance for gaining a detailed understanding of their behavior. 
Examples include responsive hydrogels used in targeted drug delivery, responsive surfaces, or responsive valves in microfluidic systems.
A detailed understanding of the mechanical behavior of such responsive materials is also highly relevant to the design of materials in the promising field of shape-morphing materials, where mechanical instabilities are exploited to achieve highly controlled, complex shapes based on relatively simple structures \citep{SydneyGladman2016413}.

However, the inherent link between thermal and mechanical properties has not been exploited to extract mechanical properties directly from calorimetric measurements.
Such an approach would be especially valuable for sub-micron or oddly shaped temperature-responsive objects such as microgels, and hydrogel particles, and, potentially, for temperature-sensitive biological materials as well. For all of these materials mechanical properties are difficult to measure using traditional techniques. 

In this paper, we show that compressive elastic moduli can be probed using DSC measurements, without directly measuring any forces or stresses. 
To do so, we exploit the inherent thermodynamic link between calorimetric and mechanical material properties of temperature-sensitive hydrogels.
We use macroscopic PNIPAM hydrogels as a model material to test our calorimetry-based measurement method. To validate our approach, we compare the results to traditional osmotic compression measurements on the same samples, finding good agreement. 

\section{Materials and methods}
\label{materialsandmethods}

\textbf{PNIPAM hydrogels} are synthesized by mixing 40 mL of Milli-Q water (resistivity$> 18\,\mathrm{M}\Omega \mathrm{cm}$) with 5 wt\% NIPAM (N-isopropylacrylamide, Sigma Aldrich), 0.25 wt\% of the cross-linker (Methylene bisacrylamide, Sigma Aldrich ) and 0.1 wt\% of the photo-initiator (Irgacure, Sigma Aldrich), corresponding to a monomer-to-cross-linker weight ratio of 100:5. 
After shaking, we transfer the solution to a well plate (2 mL, VWR) and place it under a UV lamp (Vilber-Lourmat, model VL215.LC, 60W), distance of approximately 9 cm) for 1h, which results in the formation of a cross-linked PNIPAM gel network. 
The resulting hydrogels are cut into a few pieces, which constitute the individual samples we work with in all further experiments. 
These samples are then placed in deionized water and allowed to swell overnight. 
The equilibrium volume of each piece of hydrogel is determined by removing the individual sample from the water bath, removing any extra water from its surface, and weighing it on a balance. 
In calculating the sample volume from this weight, we assume the sample's density to be equal to that of water.  \newline

\textbf{DSC measurements}, are performed on single, millimeter-sized, pieces of hydrogel in different states of hydration. 
We prepare these samples by adding different amounts of deionized water to a single dried piece of hydrogel, which subsequently swells up to incorporate all added water. 
This swelling process takes up to 30s. 
In each series of measurements, we thus use a single piece of hydrogel to scan a range of different concentrations, from highly compressed to fully swollen. 
The thermodynamic responses are studied using a heat flux differential scanning calorimeter (Q2000, TA instruments).
The temperature is increased from 10$^\circ$C to 55$^\circ$C at a rate of 0.125$^\circ$C\slash min, in analogy to previous work on pNIPAM~\citep{Grinberg1999}.  
The temperature is held at 55$^\circ$C for 10 min.
Subsequently, the sample is cooled down to 10$^\circ$C using the same rate.
Again, the sample is held at this temperature for 10 min, after which the whole procedure is repeated once.  

To test whether our heating rates ensure that the sample remains close to its equilibrium state at each temperature, we perform measurements also at 0.08 and 0.3 $^\circ$C\slash min, (see the Supplemental Material ~\citep{Supplemental}).
The results are indeed consistent with the 0.125 $^\circ$C\slash min experiments,  particularly when comparing the total area under the heat capacity peak, rather than its shape.
It is important to note that, in our experiments, we are interested only in the total heat required to increase the sample's temperature from $T_0$ to $T_\mathrm{f}$, i.e. the area under the heat capacity peak. 
\newline

\textbf{Osmotic compression} is a well-established method for characterizing the mechanical properties of hydrogel samples~\citep{FernandezNieves:2003hh}.
We employ it here for validation of our DSC method. 
In the osmotic compression method, changes in gel volume in response to an applied external osmotic pressure $\Piext$ are measured, yielding the compressive modulus $K$ of the gel~\citep{FernandezNieves:2003hh,Brain:1996}.
To perform such measurements, we place our hydrogel samples into dialysis tubes (VWR) and submerge them in dextran solutions ($M_\mathrm{w}=70\,\mathrm{kg/mol}$, from Leuconostoc, Sigma-Aldrich), with concentrations ranging from $3.5$ to $18\,\mathrm{wt}\%$, corresponding to osmotic pressures between $2.28$ and $62.5$~kPa ~\citep{SierraMartin:2011it,FernandezNieves:2003hh,Bonnet1944}. 
We then allow the gels to equilibrate for a period of one week in the dextran solutions, after which their final equilibrium weight is measured. 
From these data, we extract the equilibrium volumes as a function of the applied osmotic pressure, $V(\Piext)$.
Because the starting weight of PNIPAM is known, the volume can be calculated based on the density of water and the PNIPAM polymer, which are 1 $\mathrm{g/cm^3}$ and 1.1 $\mathrm{g/cm^3}$, respectively. 

\section{Results and Discussion}  
\label{resultsanddiscussion}
\begin{figure}[tp!] 
    \centering
    \includegraphics[width=0.5\textwidth]{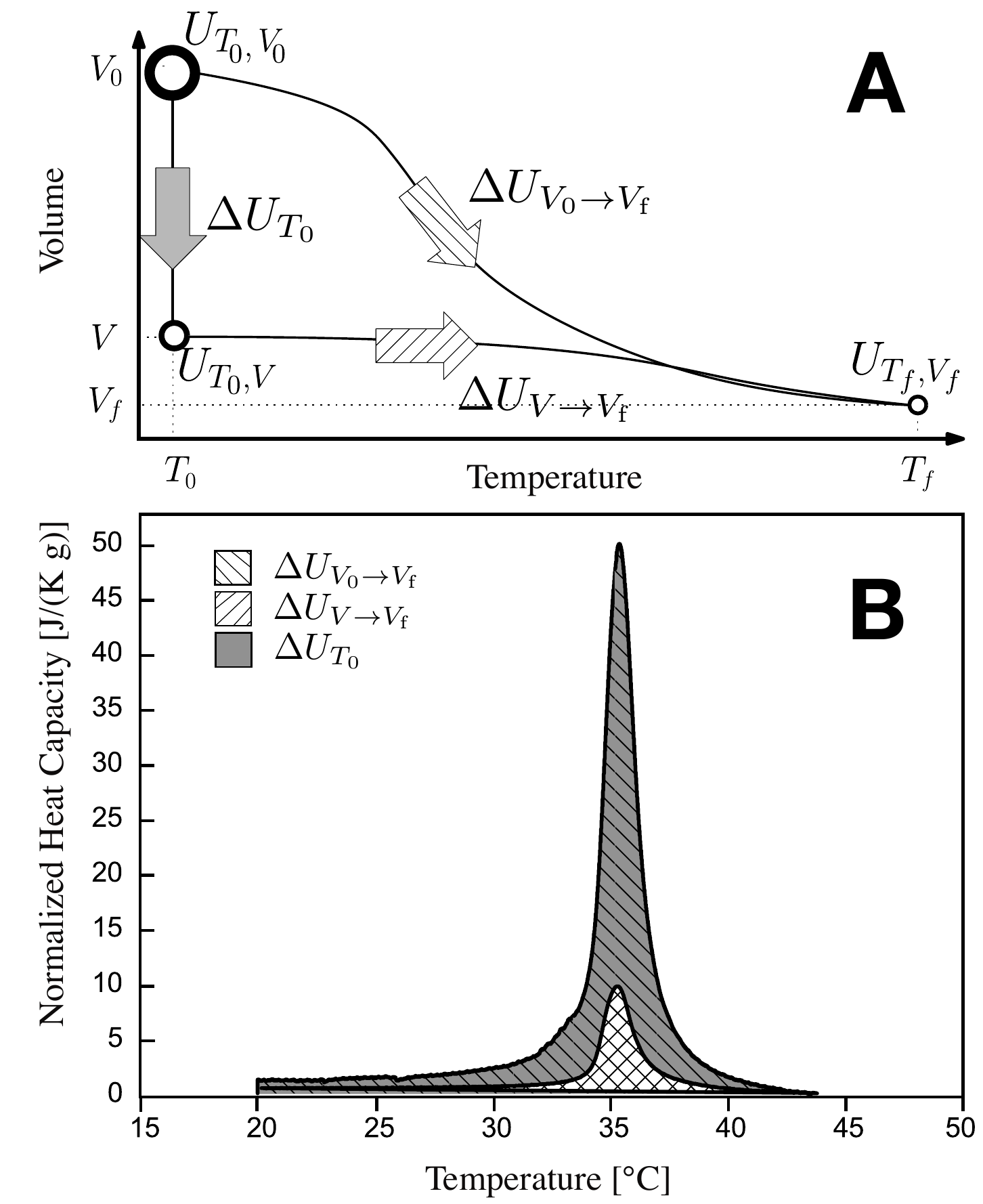}
    \caption{Extracting mechanical properties from calorimetry. 
\textbf{A)} Schematic phase diagram of calorimetric measurements. 
Energies of $\dUVo$ and $\dUV_$, for the uncompressed and compressed state, respectively, are required to heat the sample from an initial temperature $T_0$ to a final temperature $T_\mathrm{f}$. 
We assume that the energy difference $\dUT0$ is the work required to compress a particle from $V_0$ to $V$ at a constant temperature $T_0$. 
Under this assumption, a measurement of $\dUT0$ thus constitutes a mechanical measurement of the compressive modulus. 
\textbf{B)} Typical corresponding calorimetry measurements. Excess heat capacity per gram of PNIPAM as a function of temperature for (lower curve) compressed and a (higher curve) fully swollen  sample; a flat level of heat capacity, dominated by the background water, has been subtracted. 
We observe a clear peak around the LCST for both samples. 
The area under the curves corresponds to the excess energy required to heat the samples, per gram of PNIPAM. Indeed for the uncompressed sample, we find a higher area ($\backslash \backslash$-hatched) then for the compressed sample (//-hatched). 
The energy difference (the grey area) corresponds to the internal energy difference $\dUT0$, which we approximate as the mechanical work $\Wdsc$ required to compress the sample from $\Vo$ to $\Vf$.} 
    \label{fig:Simple_overview} 
\end{figure}

Our method is based on the comparison of the heat required to cross the LCST, between a fully swollen hydrogel and a gel in a compressed state. 
Since the gel does not perform work on its environment during the DSC measurement, this heat is equal to the change in internal energy. As is illustrated in Fig.~\ref{fig:Simple_overview}, the difference in internal energy between two different states of swelling at a temperature $T_0$ can be obtained as the difference in heat required to thermally deswell both samples.

Performing these experiments for a range of swelling states provides the internal energy as a function of compression at the temperature $T_0$, which we use to extract the mechanical properties of the gel at this temperature.

To compare data for gels of different water content, we first subtract the background contribution from the water. 
To do so, for each data, set we draw a line from the heat capacity value at $T$=20$^\circ$C to $T$=55$^\circ$C, which we refer to as the straight baseline. 
We subtract this baseline from the data, resulting in corrected heat capacity curves. 
By calculating the area under the corrected heat capacity data, the change in internal energy required to increase the temperature of the sample from its initial to its final temperature is determined, as shown in Fig.~\ref{fig:Simple_overview}B.
The data are further normalized by the sample dry weight so that the results for all samples can be directly compared~(see the Supplemental Material~\citep{Supplemental} for more details).

We define the normalized change in internal energy of a fully swollen piece of hydrogel, upon heating from $T_0$ to $T_\mathrm{f}$, as $\dUVo$. As shown in Fig.~\ref{fig:Simple_overview}, such a gel also undergoes a volume change from $\Vo$ to $\Vf$.
Compressing a fully swollen gel piece from $V_0$ to $V$ at constant temperature is associated with a change in internal energy $\Delta U_{T_0}$. 
Starting from this compressed state, heating the hydrogel from $T_0$ to $T_\mathrm{f}$ is associated with a change in internal energy $\dUV_$. 
In both cases, the initially fully swollen gel, at volume $V_0$ and temperature $T_0$, reaches the same final state, volume $\Vf$ at temperature $T_\mathrm{f}$; therefore, we can write $\dUVo=\Delta U_{T_0}+\dUV_$. 

In further analysis, we approximate the mechanical work required to compress the hydrogel at constant temperature as $\Wdsc \approx \dUT0$, assuming that any heat flows that would occur during an isothermal compression are small compared to the total difference in internal energy between the compressed and uncompressed states. 
A similar assumption is commonly made in macroscopic mechanical tests, where temperature changes of the sample associated with compression are generally ignored.
We also assume that the entropic contribution is relatively small because we cross-linked our hydrogels at a concentration close to the fully swollen state; entropic effects due to chain stretching are therefore not expected to play an important role.
Under these assumptions, if we measure the difference between $\dUVo$ and $\dUV_$, for different compression states, we obtain a good estimate of the associate work $W$ needed to compress the gel.
This estimated work yields a direct measure of the material's compressive elastic modulus, $K =-V\frac{\partial^2 W}{\partial V^2}$.

Calculating $K$ involves taking a double derivative; therefore, in order to avoid the amplification of experimental errors, we wish to describe the DSC data using a simple and continuous functional form. 
To arrive at a physically motivated expression, we take the integral of a pressure function, $W= \int_{V_{0}}^{V}\Piext(\tilde{V})\,d\tilde{V}$, where $\Piext(V)$ is the applied pressure as a function of volume $V$. 
To approximate this pressure function we follow the standard scaling arguments for the equilibrium swelling of gels ~\citep{rubinsteinbook}, equating the osmotic pressure difference between a semidilute polymer solution and the external pressure, $\Piext(V)$, with the elastic modulus of a gel as

\begin{equation} 
\Piext=\beta\left(\frac{V}{V_0}\right)^{-9/4}-\beta\left(\frac{V}{V_0}\right)^{-1/3}.
\label{eq:P}
\end{equation}

Here, $V/V_0$ is the ratio between the compressed volume $V$ and the fully swollen volume $V_0$, the first term on the right-hand side denotes the osmotic pressure of a semidilute polymer solution, and the second term represents the elastic modulus according to rubber elasticity. 
We treat the prefactors as fitting parameters, and, because $V=V_0$ at $\Piext=0$, the two prefactors must be identical, leaving a single fitting parameter $\beta$. 

We obtain $W$ by integrating the applied pressure $\Piext$ over the volume as

\begin{equation}
\label{eq:IntegrationP}
W=\frac{4\beta}{5} \Big(\frac{V}{V_{0}}\Big)^{-5/4} -\frac{3\beta}{2} \Big(\frac{V}{V_{0}}\Big)^{2/3}+0.7\beta  \ ,
\end{equation}
where we have added the term $0.7 \beta$ to comply with the boundary condition $W=0$ at $V=V_0$. 
Using DSC, we determined the difference in changes of internal energy (or the work $\Wdsc$) for hydrogels of different dry weight and at different degrees of compression ($V/V_0$). 
Because the concentration of PNIPAM is known at every rehydration step, $V$ and $V_0$ can be calculated based on the density of water and PNIPAM, 1.1 $\mathrm{g/cm}^3$ and 1 $\mathrm{g/cm}^3$, respectively.   

\begin{figure}[hbpt] 
        \centering
        \includegraphics[width=0.5\textwidth]{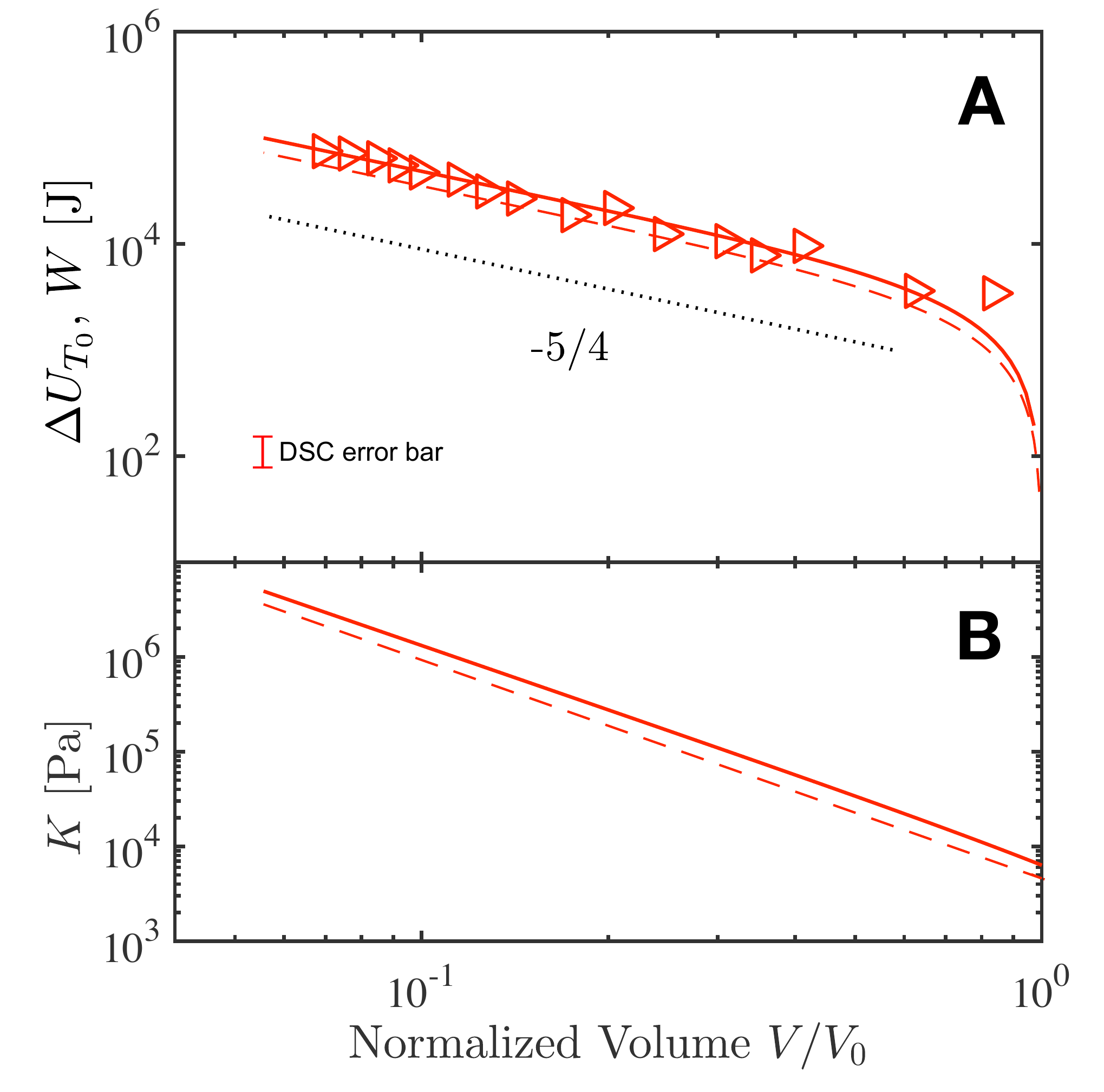}
        \caption{\textbf{A)} Work $W$ required to compress a particle from $\Vo$ to $V$ as a function of the volume ratio $V/V_0$. 
The red triangles are the calorimetric results ($\Delta U_{T_0}$) and the solid line is the fit using Eq.~(\ref{eq:IntegrationP}), resulting in $\beta \approx 3.29$~kPa. 
This fit is in good agreement with the prediction based on the osmotic compression measurement ($\Wosm$), plotted as a dashed line. 
The dotted line indicates a slope of $-5/4$ which is the dominant term in Eq.~(\ref{eq:IntegrationP}).
The average error for DSC experiments is 10 \%, as indicated by the error bar in the bottom-left corner.  
\textbf{B)} Corresponding compressive modulus $K$ as a function of $V/V_0$. At $V=V_0$, we find $K=6.32 \pm 1.43$~kPa and $4.59 \pm 0.52$~kPa based on DSC (the solid line) and osmotic compression (the dashed line), respectively.
        } 
        \label{fig:Energy_stiffness}
\end{figure}

The experimental DSC results are plotted as red triangles in Fig.~\ref{fig:Energy_stiffness}A and the corresponding fit to the experimental data is shown as a solid line. 
We see that the experimental data clearly follows the dominant power-law term with exponent $-5/4$ from Eq.~(\ref{eq:IntegrationP}), within most of the fitting range, as indicated by the dotted line. 
This agreement demonstrates that our choice for the functional form of $\Piext$ is appropriate.
Thus, in our approach, the compressive modulus $K$ can be determined directly from $\Wdsc$, via a double derivative, as $K(V)= -V \frac{\partial \Piext}{\partial V} =-V\frac{\partial^2 \Wdsc}{\partial V^2}$, shown in Fig.~\ref{fig:Energy_stiffness}B as the solid line. 
In the swollen state, at $V=V_0$, we obtain $K\approx~6.3~\pm~1.43~\mathrm{kPa}$. 

It is important to note that, in our experiments, we are interested only in the total energy required to heat the sample from $T_0$ to $T_\mathrm{f}$, i.e., the area under the heat capacity curve.  
Thus, as long as we compare experiments performed at the same heating rate, any rate-dependent smearing of the heat capacity peak would be relatively inconsequential.
We show in the Supplemental Material~\citep{Supplemental} that, for heating rates of 0.08 and 0.3 $^\circ$C\slash min, we indeed obtain a similar result as for the 0.125 $^\circ$C\slash min experiments.

\begin{figure}[hbpt] 
	\centering
	\includegraphics[width=0.47\textwidth]{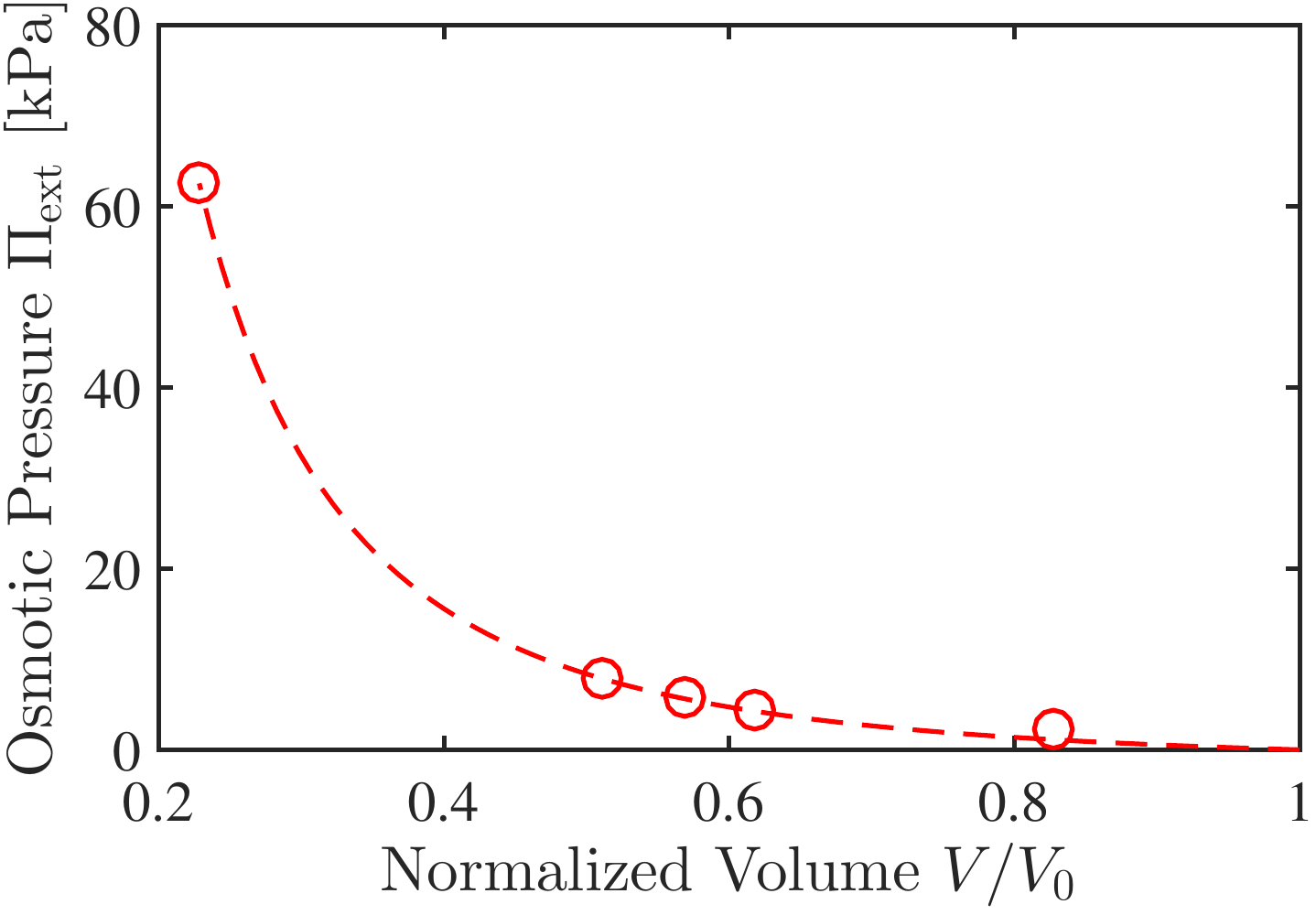}
	\caption{
Osmotic pressure as a function of the normalized hydrogel volume $V/V_0$. 
The red circles are the experimental data from the osmotic compression tests and the dashed line is a fit to Eq.(\ref{eq:P}), with $\beta$=2.39 $\pm$ 0.06~kPa. 
The error bars are smaller than 1\%, which is smaller than the marker size; see the error analysis in the Supplemental Material ~\citep{Supplemental}.}
	\label{fig:Osmotic}
\end{figure}

To verify our DSC-based method, we perform conventional osmotic compression experiments. 
By doing so, we can verify our chosen pressure function, the resulting energy function, and our assumption that $\Wdsc \approx \Delta U_{T_0}$.
The experimental results are plotted as open circles in Fig.~\ref{fig:Osmotic} and the fit using Eq.~(\ref{eq:P}) is depicted as the dashed line.
The volume-dependent compressive modulus $K(V)$ can be determined from its definition, $K\left( V \right)= - \left( V/V_0 \right) \frac{\partial \Piext}{\partial (V/V_0)}$, and,  finally $\Wosm$ can be determined by applying Eq.~(\ref{eq:IntegrationP}).
The results are shown as the dashed lines in Fig.~\ref{fig:Energy_stiffness}A and \ref{fig:Energy_stiffness}B. 
Using this approximation for the pressure function, we extract a compressive modulus as a function of $V/V_0$. 
At the equilibrium volume, we obtain $K(V_0)\approx 4.6 
\pm 0.52\ \mathrm{kPa}$, in good agreement with the $K\approx~6.3~\pm~1.32~\mathrm{kPa}$ value obtained via the DSC method. The uncertainties associated with both methods are estimated in a separate error analysis, included in the Supplemental Material~\citep{Supplemental}.

The comparable results for $\Wdsc$ and $\Wosm$ clearly indicate that the two completely different experimental approaches lead to consistent results, thereby confirming the validity of our DSC-based approach and our assumption of $\Wdsc \approx \dUT0$. 
The good agreement between the two measurements thus indicates that, for these materials, the mechanical work needed to compress a particle can be approximated by the difference in internal energy between the swollen and compressed states, neglecting any heat flows that occur during compression. 

However, it is not obvious that this approximation is justified for all types of temperature-sensitive gel materials. 
The role of entropy is expected to be largest for gels that are cross-linked in the collapsed state, as swelling of such gels requires significant stretching of polymer chains. 
However, even in this case, enthalpic effects should play a major role and constitute a significant fraction of the free energy difference between the swollen and compressed  states of the gel.  

In our experiments, the gels were cross-linked at room temperature, at a concentration near the fully swollen state of the system.  
For this case, compressing the material results in much smaller changes in chain entropy, as the cross-linking points are brought closer together with respect to the equilibrium chain configuration, with no associated stretching of chains.  
Moreover, compression will result in a reduction of both the mixing entropy and the configurational chain entropy. 
As a result, for this case the DSC-based measurement in fact represents a \emph{lower bound} on the compressive modulus. 
While we expect our approximation to be justified for various types of gels and other $T$-sensitive materials, its validity should be checked by comparison to separate mechanical measurements, as we do for the PNIPAM gels used here. 

\section{Conclusions}
\label{conclusions}

In this paper, we perform calorimetric measurements on PNIPAM hydrogels at different levels of compression and show that the change in internal energy required to heat a sample at starting temperature $T_0$ to a final temperature $T_\mathrm{f}$ above its LCST is lower than that for an uncompressed sample that undergoes the same heat treatment. 
The difference between these two changes in internal energy must be equal to the difference in internal energy between the initially compressed and the uncompressed states, respectively. 
We approximate this difference in internal energy as the work $W$ required to compress the gels at a constant temperature $T_0$, which in turn yields the compressive (bulk) modulus $K$. 
The most direct justification for the validity of this approximation is the good agreement we obtain between the conventional osmotic compression measurements and the DSC-based measurements, as shown in Fig. \ref{fig:Energy_stiffness}.

To minimize the amplification of experimental errors introduced by the derivatives involved, we fit the DSC and osmotic compression data to a physically meaningful functional form and use this approximation to extract elastic moduli.

We note that to achieve accurate measurements of the compressive modulus, a series of DSC experiments across a significant range of concentrations is required.
Moreover, the heating rate in the DSC experiment should be slow enough to give the gels sufficient time to remain close to their equilibrium state during (de)swelling. 
As a result, the measurements are relatively time consuming.

However, the sample preparation for the DSC experiments is straightforward and we expect the method to be applicable to a wide range of materials that exhibit a pronounced volume phase transition.
Our results demonstrate that calorimetry can be a powerful technique to probe the mechanical properties of temperature-responsive gels.  
Importantly, the method does not require the measurement of any stresses or pressures, as would be required in a traditional mechanical test. 
Our method can therefore be particularly useful for sub-micron and oddly shaped thermosensitive materials for which established methods are inadequate. 

Future improvements to our approach, including a better understanding of the changes in the physical state of thermosensitive materials during heating, could enable detailed measurements of the compressive modulus as a function of temperature for a wide range of materials. 
We expect our approach to be immediately useful for the study of temperature-sensitive soft materials, such as those used in smart drug delivery systems, shape-morphing materials, and responsive soft actuators.

We thank Paul van der Schoot for the valuable discussions. The work of F.J.A. and H.M.W. forms part of the research program of the Dutch Polymer Institute (DPI), Project No. 738; we are grateful for its financial support. J.M. and H.M.W. are also grateful to the Royal Society (Grant No. IE111253) for the financial support.

\clearpage
\section*{SUPPLEMENTARY INFORMATION}
Literature references below refer to the reference section of the main paper above.
\appendix

\def\Vf{ {V_{\mathrm{f}}} }
\def\Vo{V_{\mathrm{0}}}
\def\Piext{\Pi_\mathrm{ext}}
\def\Wdsc{W_{\mathrm{dsc}}}
\def\Wosm{W_\mathrm{osm}}

\begin{center}\rule{3in}{0.4pt}\end{center}
\section{DSC data analysis}
\begin{figure*}[p]
    \centering
    \includegraphics[width=0.9\linewidth]{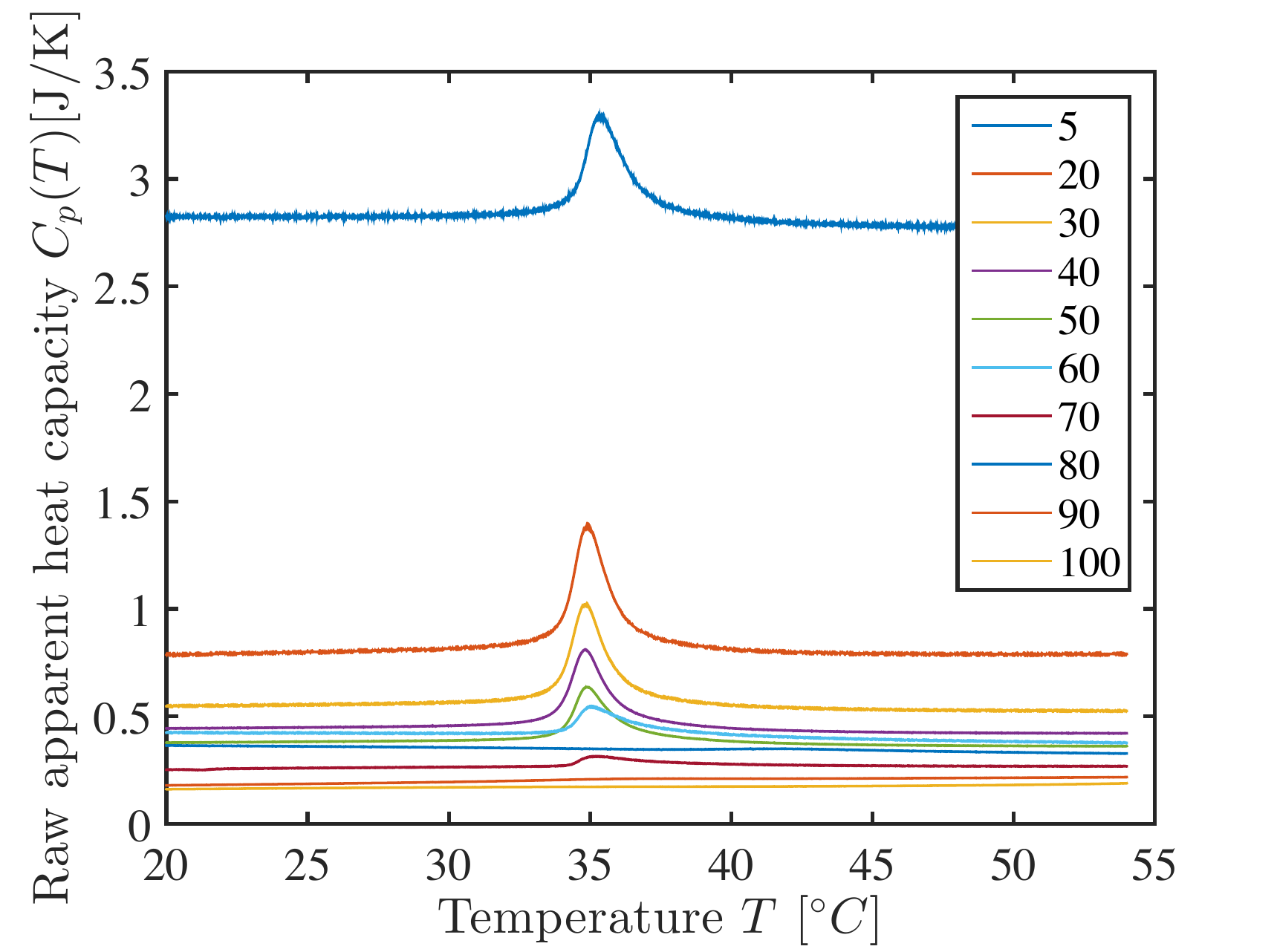}
    \caption{Raw apparent heat capacity $C_\mathrm{p}(T)$ as a function of temperature for samples with different states of compression and pNIPAM dry weights as listed in Table~1. The legend indicates the pNIPAM concentration in wt\%.}
    \label{fig:raw}
\end{figure*}

Differential scanning calorimetry (DSC) measurements are performed on pNIPAM
hydrogels prepared from rehydration of dried gels, as described in Section II.
We study hydrogels from different batches, each with a different pNIPAM dry
weight. All measurements are performed at a rate of 0.125
$^\circ\mathrm{C}/\mathrm{min}$ within a temperature range from $T=$ 10--55
$^\circ C$.  The determined differential heat flux is converted to an apparent
heat capacity $C_p(T)$ through division by the imposed heating rate.  The
apparent heat capacity $C_p(T)$ is plotted for gels prepared in different
states of hydration in Fig.\ref{fig:raw}.  The corresponding  states of
compression (in terms of the volume ratio $V/V_0$) and dry weights
$m_{\mathrm{pNIPAM}}$ are summarized in Table \ref{Tablesummary}. 

\setlength{\tabcolsep}{10pt}
\begin{table*}[t]
	\centering
	\caption{Volume ratios and dry weights for samples and curves shown in Fig.\ref{fig:raw}}
	\label{Tablesummary}
	\begin{tabular}{c|c|c}
		Concentration [wt\%] & Volume ratio $V/V_0$ & $m_\mathrm{pNIPAM}$ [mg] \\\hline
		5                & 1	     & 26.5   	\\
		20               & 0.25      & 34.6   	\\
		30               & 1/6       & 28.7   	\\
		40               & 1/8       & 28.7   	\\
		50               & 1/10      & 28.7   	\\   
		60               & 1/12      & 28.5   	\\     
		70               & 1/14      & 34.3   	\\     
		80               & 1/16      & 28.5   	\\     
		90               & 1/18      & 34.6   	\\	
		100              & 1/20      & 35.3  	\\
				
	\end{tabular}
\end{table*}

In the temperature range of interest, the heat capacity of water is featureless, and since gels at different degrees of hydration contain different amounts of water, the total measured apparent heat capacity will be shifted along the $y$-axis, as observed in Fig.\ref{fig:raw}.

\begin{figure*}[p]
	\centering
	\includegraphics[width=0.9\linewidth]{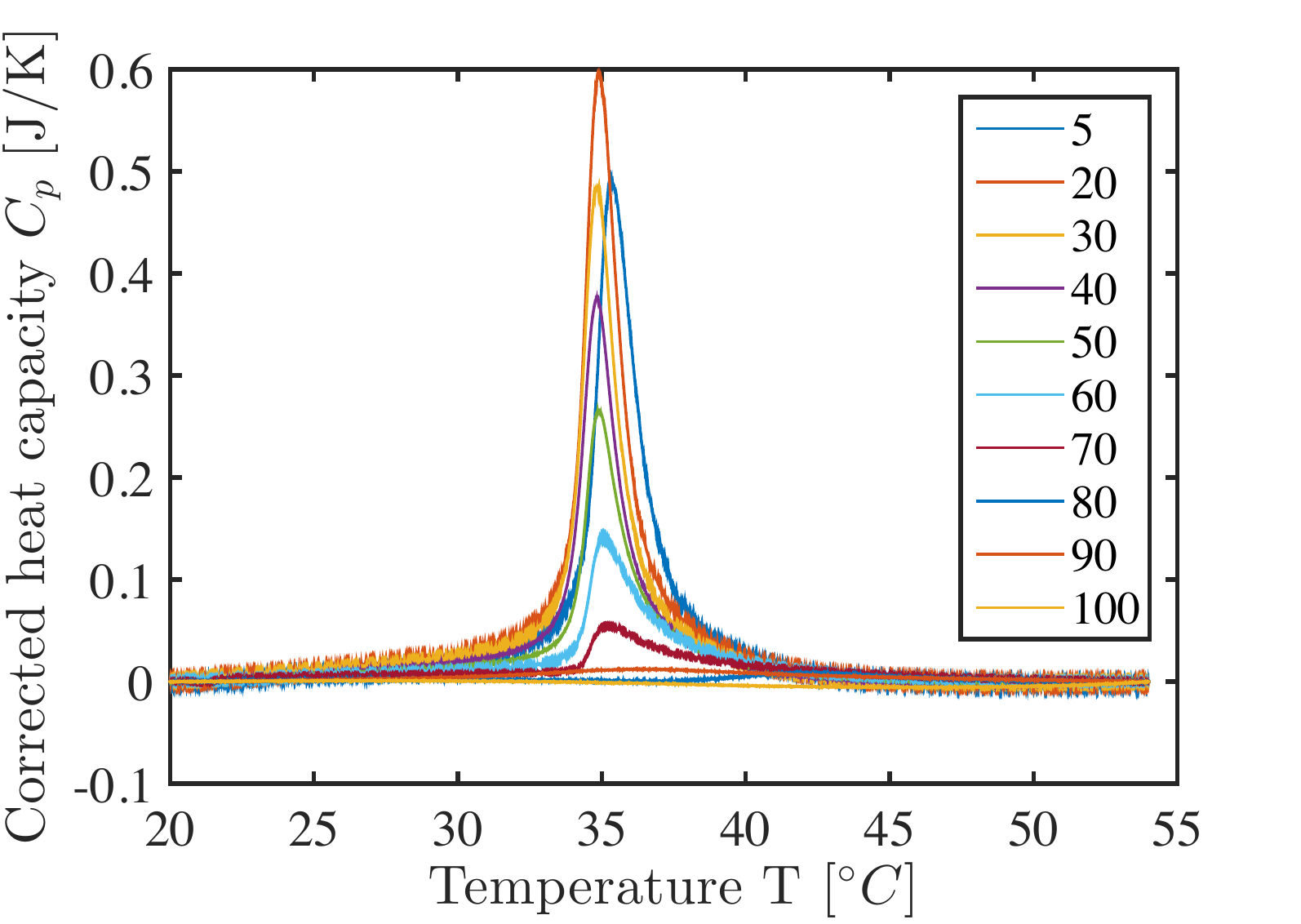}
	\caption{Corrected heat capacity as function of temperature for samples with different compression rates and pNIPAM dry weight.  }
	\label{fig:corrected}
\end{figure*}

Thus, to compare data for gels of different water content, we subtract this contribution of to the background water. 
To do so, for each data set we draw a straight line from heat the capacity value at $T$=20$^\circ C$ to that at $T$=55$^\circ C$; we refer to this line as the straight baseline. 
We subtract this baseline from the data, the result is show in Fig. \ref{fig:corrected}.

\begin{figure*}[p]
	\centering
	\includegraphics[width=0.9\linewidth]{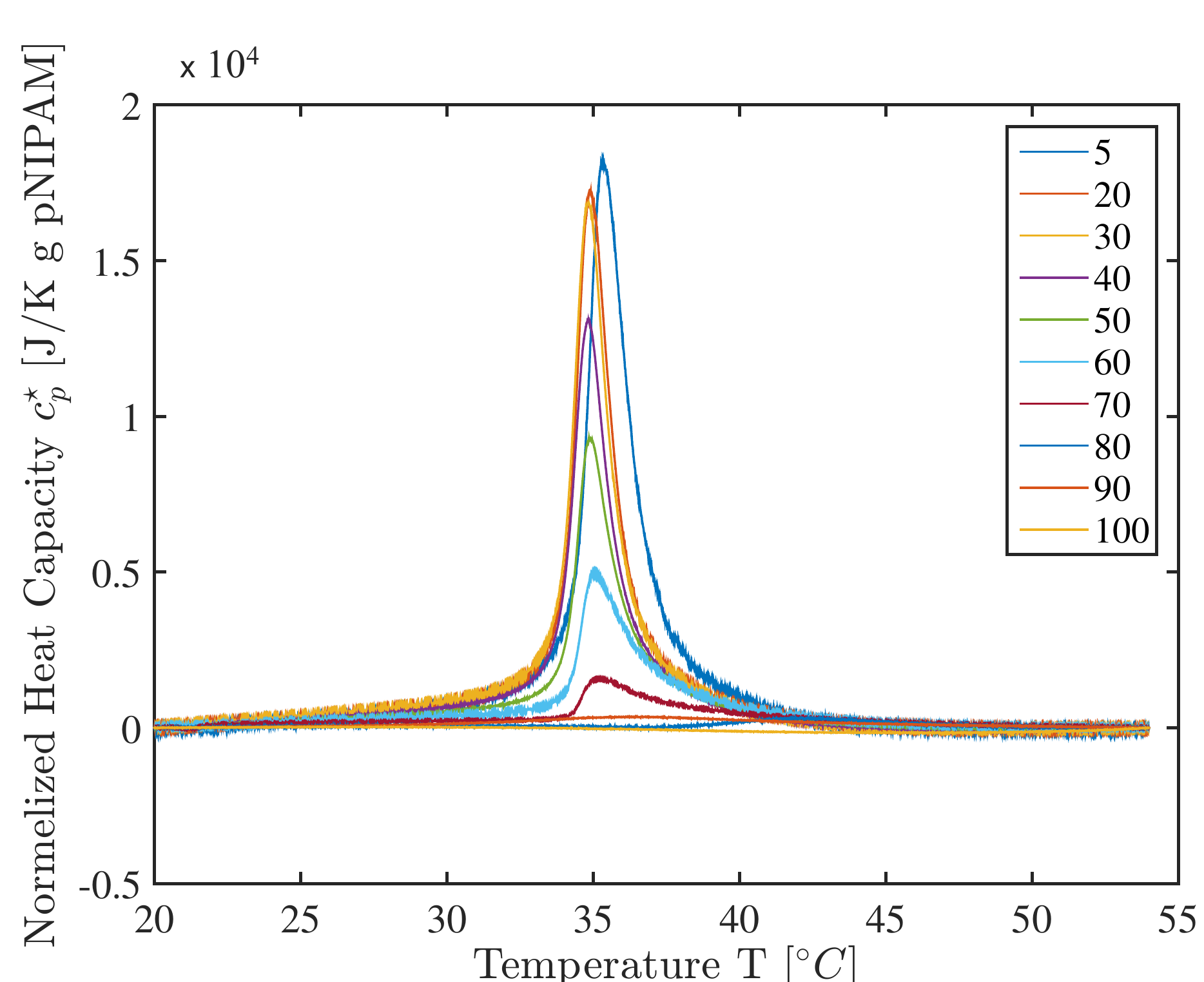}
	\caption{Heat capacity normalized with the dry weight of pNIPAM as a function of temperature. 
		The top curve is for the uncompressed particle and the lower curves are for the pNIPAM gel pieces in different states of compression. 
		The difference in area between the top curve and the lower curves is $\Delta U_{T_0}$, which we set equal to the mechanical work $\Wdsc$ performed on the material when compressing it from an initial volume $V_0$ to a final volume $V$.}
	\label{fig:normalised}
\end{figure*}

Moreover, to account for the variation in the pNIPAM dry weight, we calculate the specific heat capacity, $c_p^*(T)$, by normalizing the $C_p^*(T)$ measured for each sample with its dry weight. 
The resultant specific apparent heat capacities $c_p^*(T)$ are plotted in Fig.\ref{fig:normalised}, where the uppermost data represents the fully swollen hydrogel and the area under this curve is $\Delta U_{V_0 \rightarrow V_f}$. 
By subtracting the data obtained for each degree of hydration (compression) from the data for the fully swollen state, we obtain $\Delta U_{T_0}$, the difference in internal energy between the swollen state ($T_0$, $V_0$) and the compressed state ($T_0$, $V$). 

To extract the elastic properties of the sample from these data, we then approximate $W_\mathrm{dsc} \approx \Delta U_{T_0}$, where $W_\mathrm{dsc}$ is the mechanical work that would be required for the isothermal compression of a particle at temperature $T_0$ from an initial volume $V_0$ to a compressed volume $V$.
From this mechanical work $W_\mathrm{dsc}$ as a function of $V$ the compressive modulus of the gels can be determined directly, as discussed in more detail in the main manuscript. 

\section{Influence of heating rate}
\begin{figure*}[p]
	\centering
	\includegraphics[width=0.9\linewidth]{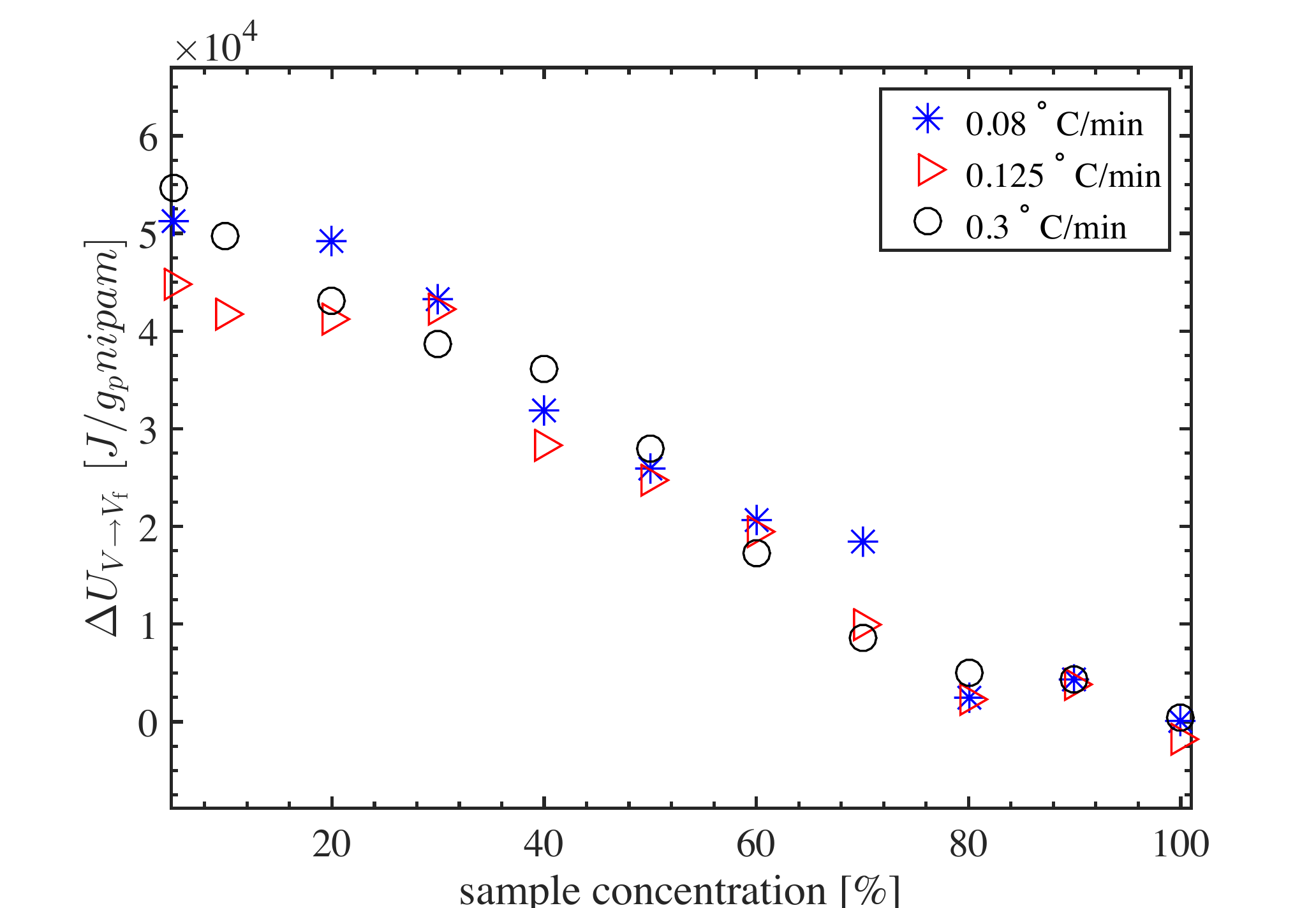}
	\caption{$\Delta U_{V \rightarrow V_\mathrm{f}}$ as function of volume fraction of pNIPAM for 3 different heating rates. 
The blue cross is 0.08 $^\circ\mathrm{C}/\mathrm{min}$, the red triangles is 0.125 $^\circ\mathrm{C}/\mathrm{min}$ and the black circles is 0.3 $^\circ\mathrm{C}/\mathrm{min}$. 
The heating rate appears to only have a limited effect on the measured values of $\Delta U_{V \rightarrow V_\mathrm{f}}$. }
	\label{fig:heatrate}
\end{figure*}

We also investigated the influence of the heating rate on the $\Delta U_{V \rightarrow V_\mathrm{f}}$. We performed experiments on the same samples as in table \ref{Tablesummary} but now at 0.08 $^\circ\mathrm{C}/\mathrm{min}$ and 0.3 $^\circ\mathrm{C}/\mathrm{min}$ and compared these to the 0.125 $^\circ\mathrm{C}/\mathrm{min}$ measurements. As can been seen in Fig. \ref{fig:heatrate} this seems not to have a significant influence, however faster heating rates where not tried and it must be noted that this could influence $\Delta U_{T_0}$. 

\section{Error Analysis}

\subsection{Osmotic Compression Experiments}
We apply an osmotic compression by placing the pNIPAM particles in a dialysis bag and submerging it in a dextran solution. 
We vary the dextran weight concentration $c$ between 3.5 and 18 wt\%. The corresponding applied osmotic pressures $\Pi$ are calculated, following Bonnet-Gonnet et al. \cite{Bonnet1944}, as

\begin{equation}
\Pi=286c+87c^2+5c^3 \, ,
\end{equation}

where $c$ is the polymer concentration in wt\% and $\Pi$ is the osmotic pressure in Pa.
Using this functional form for the dependence of pressure on concentration, we thus estimate the relative error in the applied pressure $\Delta \Pi / \Pi$ as
\begin{eqnarray}
\frac{\Delta \Pi}{\Pi}=(\frac{\partial \Pi}{\partial c} \Delta c)/\Pi,\\
\frac{\Delta \Pi}{\Pi}=\frac{\big( 286+87c+15c^2\big) \Delta c}{286c+87c^2+5c^3},\\
\frac{\Delta \Pi}{\Pi}=\frac{\big( 286+87c+15c^2\big) c}{286c+87c^2+5c^3} \frac{\Delta c}{c}
\end{eqnarray}
where $\Delta c / c$ is the error in measuring the weight of the dextran; 
we estimate this as $\Delta c / c \approx 1\% $ for the scale we have used. 

This results in a error between the 0.1\% and 0.35 \% in the concentration range of 3.5 to 18 wt\%. 
This error is smaller than the marker size used in Fig. 3 in the main text. As a result, no visual error bars are included in this figure.  
 
\section{Compressive Modulus Calculations}
\subsection{osmotic compression experiment}
We calculate the compressive (bulk) modulus based on the osmotic compression experiment as
\begin{equation}
K=-\frac{\partial \Pi}{\partial V} V,
\end{equation}
where $\Pi$ is the osmotic pressure, which we assume follows the functional form
\begin{equation}
\Pi=\beta\left(\frac{V}{V_0}\right)^{-9/4}-\beta\left(\frac{V}{V_0}\right)^{-1/3} \, .
\end{equation}

The associated relative error in the modulus can thus be estimated as
\begin{equation}
\frac{\Delta K}{K}=\Big(\frac{\partial K}{\partial V} \Delta V +\frac{\partial K}{\partial \Pi} \Delta \Pi \Big)/K.
\end{equation}

If for the purposes of this error analysis we only consider the dominant term ($V^{-9/4}$) this results in
\begin{eqnarray*}
\frac{\Delta K}{K}=\frac{\frac{\partial}{\partial V}\big( V^{-9/4}\big)\Delta V}{V^{-9/4}}+\frac{\Delta P}{P},\\
\frac{\Delta K}{K}=\frac{\frac{-9}{4} V^{-13/4} \Delta V}{V^{-9/4}}+0\\
\frac{\Delta K}{K}=9/4 \frac{\Delta V}{V}.
\end{eqnarray*}
Assuming a relative error for the volume of $\Delta V/V$=0.05, we obtain for the modulus a relative error of $\frac{\Delta K}{K} \approx 11.25 \%$.
As a result, at $V/V_0=1$ we obtain $K=4.6 \pm 0.52$ kPa for the osmotic compression experiment. 

\subsection{DSC experiment}
We calculate the compressive modulus based on the DSC experiments as
\begin{equation}
K=-\frac{\partial^2 W}{\partial V^2} V.
\end{equation}
With $W$ being
\begin{equation}
W=\frac{4\beta}{5} \Big(\frac{V}{V_{0}}\Big)^{-5/4} -\frac{3\beta}{2} \Big(\frac{V}{V_{0}}\Big)^{2/3}+0.7\beta  \ ,
\end{equation}

the relative error we make in this is
\begin{equation}
\frac{\Delta K}{K}=\left( \frac{\partial K}{\partial V} \Delta V +\frac{\partial K}{\partial W} \Delta W \right)\frac{1}{K} \, .
\end{equation}

If for the purposes of this error analysis we only consider the dominant term ($V^{-5/4}$) this results in 

\begin{eqnarray*}
\frac{\Delta K}{K}=\frac{\Big( \frac{\partial^2 W}{\partial V^2} +V\frac{\partial^3 W}{\partial V^3}\Big) \Delta V }{V\frac{\partial^2 W}{\partial V^2}}+\frac{\Delta W}{W}, \\
\frac{\Delta K}{K}=\frac{\Delta V}{V}+\frac{\frac{\partial^3 W}{\partial V^3}\Delta V}{\frac{\partial^2 W}{\partial W^2}}+\frac{\Delta W}{W}, \\
\frac{\Delta K}{K}=\frac{\Delta V}{V}+\frac{13*9/(4*4)*V^{-17/4}\Delta V}{9/4*V^{-9/4}}+\frac{\Delta W}{W}, \\
\frac{\Delta K}{K}=(1+13/4)\frac{\Delta V}{V}+\frac{\Delta W}{W}, \\
\frac{\Delta K}{K}=4.25\frac{\Delta V}{V}+\frac{\Delta W}{W}. \\
\end{eqnarray*}
If we assume $\Delta V/V$=0.05 and $\frac{\Delta W}{W}=0.1$ we obtain
$\frac{\Delta K}{K} \approx 31.25 \%$.
Therefore at $V/V_0=1$ we obtain $K=6.3\pm 1.43$ kPa from the DSC experiment. 
Including these errors, the values of $K_{\mathrm{dsc}}=6.3\pm 1.43$ and $K_{\mathrm{osm}}=4.6 \pm 0.52$ overlap, thus indicating a validation of our DSC-based approach by conventional osmotic compression measurements.


\begin{thebibliography}{24}
    

\expandafter\ifx\csname natexlab\endcsname\relax\def\natexlab#1{#1}\fi
\expandafter\ifx\csname bibnamefont\endcsname\relax
  \def\bibnamefont#1{#1}\fi
\expandafter\ifx\csname bibfnamefont\endcsname\relax
  \def\bibfnamefont#1{#1}\fi
\expandafter\ifx\csname citenamefont\endcsname\relax
  \def\citenamefont#1{#1}\fi
\expandafter\ifx\csname url\endcsname\relax
  \def\url#1{\texttt{#1}}\fi
\expandafter\ifx\csname urlprefix\endcsname\relax\def\urlprefix{URL }\fi
\providecommand{\bibinfo}[2]{#2}
\providecommand{\eprint}[2][]{\url{#2}}


\bibitem[{\citenamefont{Kim and Lee}(1999)}]{Kim1999} 
\bibinfo{author}{\bibfnamefont{S.Y.} \bibnamefont{Kim}} \bibnamefont{and}
  \bibinfo{author}{\bibfnamefont{Y.M.} \bibnamefont{Lee}},
  \bibinfo{title}{Drug release behavior of electrical responsive poly(vinyl alcohol)/poly(acrylic acid) IPN hydrogels under an electric stimulus},
  \bibinfo{journal}{Journal of Applied Polymer Science}
  \textbf{\bibinfo{volume}{74}}, \bibinfo{pages}{1752} (\bibinfo{year}{1999}).

\bibitem[{\citenamefont{Wu et~al.}(2003)\citenamefont{Wu, Zhou, and
  Hu}}]{Wu:2003ue}  
\bibinfo{author}{\bibfnamefont{J.}~\bibnamefont{Wu}},
  \bibinfo{author}{\bibfnamefont{B.}~\bibnamefont{Zhou}}, \bibnamefont{and}
  \bibinfo{author}{\bibfnamefont{Z.}~\bibnamefont{Hu}},
  \bibinfo{title}{Phase Behavior of Thermally Responsive Microgel Colloids},
  \bibinfo{journal}{Physical Review Letters} \textbf{\bibinfo{volume}{90}},
  \bibinfo{pages}{48304} (\bibinfo{year}{2003}).

\bibitem[{\citenamefont{Saunders and Vincent}(1996)}]{Saunders:1996vt}
\bibinfo{author}{\bibfnamefont{B.R.} \bibnamefont{Saunders}} 
\bibnamefont{and}
  \bibinfo{author}{\bibfnamefont{B.}~\bibnamefont{Vincent}},
  \bibinfo{title}{Thermal and osmotic deswelling of poly (NIPAM) microgel particles},
  \bibinfo{journal}{J. Chem. Soc. Faraday Trans.}
  \textbf{\bibinfo{volume}{92}}, \bibinfo{pages}{3385} (\bibinfo{year}{1996}).

\bibitem[{\citenamefont{Arleth et~al.}(2005)\citenamefont{Arleth, Xia, Hjelm,
  Wu, and Hu}}]{Arleth:2005} 
\bibinfo{author}{\bibfnamefont{L.}~\bibnamefont{Arleth}},
  \bibinfo{author}{\bibfnamefont{X.H.} \bibnamefont{Xia}},
  \bibinfo{author}{\bibfnamefont{R.P.} \bibnamefont{Hjelm}},
  \bibinfo{author}{\bibfnamefont{J.Z.} \bibnamefont{Wu}}, \bibnamefont{and}
  \bibinfo{author}{\bibfnamefont{Z.B.} \bibnamefont{Hu}},
  \bibinfo{title}{Volume transition and internal structures of small poly(N-isopropylacrylamide) microgels},
  \bibinfo{journal}{Journal of Polymer Science Part B-Polymer Physics}
  \textbf{\bibinfo{volume}{43}}, \bibinfo{pages}{849} (\bibinfo{year}{2005}).
  
  \bibitem[{\citenamefont{Yamazaki et~al.}(1999)\citenamefont{Yamazaki, Winnik,
        Cornelius, and Brash}}]{Yamazaki1999103} 
  \bibinfo{author}{\bibfnamefont{A.}~\bibnamefont{Yamazaki}},
  \bibinfo{author}{\bibfnamefont{F.}~\bibnamefont{Winnik}},
  \bibinfo{author}{\bibfnamefont{R.}~\bibnamefont{Cornelius}},
  \bibnamefont{and} \bibinfo{author}{\bibfnamefont{J.}~\bibnamefont{Brash}},
  \bibinfo{title}{Modification of liposomes with N-substituted polyacrylamides: Identification of proteins adsorbed from plasma},
  \bibinfo{journal}{Biochimica et Biophysica Acta - Biomembranes}
  \textbf{\bibinfo{volume}{1421}}, \bibinfo{pages}{103} (\bibinfo{year}{1999}).

\bibitem[{\citenamefont{Liu et~al.}(2010)\citenamefont{Liu, Ye, Yu, Liang, Liu,
  and Zhou}}]{Liu201012377} 
\bibinfo{author}{\bibfnamefont{X.}~\bibnamefont{Liu}},
  \bibinfo{author}{\bibfnamefont{Q.}~\bibnamefont{Ye}},
  \bibinfo{author}{\bibfnamefont{B.}~\bibnamefont{Yu}},
  \bibinfo{author}{\bibfnamefont{Y.}~\bibnamefont{Liang}},
  \bibinfo{author}{\bibfnamefont{W.}~\bibnamefont{Liu}}, \bibnamefont{and}
  \bibinfo{author}{\bibfnamefont{F.}~\bibnamefont{Zhou}},
  \bibinfo{title}{Switching water droplet adhesion using responsive polymer brushes},
\bibinfo{journal}{Langmuir} \textbf{\bibinfo{volume}{26}},
  \bibinfo{pages}{12377} (\bibinfo{year}{2010}).

\bibitem[{\citenamefont{Zhu et~al.}(2012)\citenamefont{Zhu, Lu, Peng, Chen, and Yu}}]{Zhu20124017} 
\bibinfo{author}{\bibfnamefont{C.-H.} \bibnamefont{Zhu}},
\bibinfo{author}{\bibfnamefont{Y.}~\bibnamefont{Lu}},
\bibinfo{author}{\bibfnamefont{J.}~\bibnamefont{Peng}},
\bibinfo{author}{\bibfnamefont{J.-F.} \bibnamefont{Chen}}, \bibnamefont{and}
\bibinfo{author}{\bibfnamefont{S.-H.} \bibnamefont{Yu}},
\bibinfo{title}{Photothermally sensitive poly(N-isopropylacrylamide)/graphene oxide nanocomposite hydrogels as remote light-controlled liquid microvalves},
\bibinfo{journal}{Advanced Functional Materials}
\textbf{\bibinfo{volume}{22}}, \bibinfo{pages}{4017} (\bibinfo{year}{2012}).

\bibitem[{\citenamefont{Yan et~al.}(2015)\citenamefont{Yan, Huang, Zhang, and
  Zhou}}]{APP:APP41669} 
\bibinfo{author}{\bibfnamefont{Y.}~\bibnamefont{Yan}},
  \bibinfo{author}{\bibfnamefont{L.}~\bibnamefont{Huang}},
  \bibinfo{author}{\bibfnamefont{Q.}~\bibnamefont{Zhang}}, \bibnamefont{and}
  \bibinfo{author}{\bibfnamefont{H.}~\bibnamefont{Zhou}},
  \bibinfo{title}{Concentration effect on aggregation and dissolution behavior of poly(N-isopropylacrylamide) in water},
  \bibinfo{journal}{Journal of Applied Polymer Science}
  \textbf{\bibinfo{volume}{132}} (\bibinfo{year}{2015}).

\bibitem[{\citenamefont{Afroze et~al.}(2000)\citenamefont{Afroze, Nies, and
  Berghmans}}]{Afroze200055} 
\bibinfo{author}{\bibfnamefont{F.}~\bibnamefont{Afroze}},
  \bibinfo{author}{\bibfnamefont{E.}~\bibnamefont{Nies}}, \bibnamefont{and}
  \bibinfo{author}{\bibfnamefont{H.}~\bibnamefont{Berghmans}},
  \bibinfo{title}{Phase transitions in the system poly(N-isopropylacrylamide)/water and swelling behaviour of the corresponding networks},
  \bibinfo{journal}{Journal of Molecular Structure}
  \textbf{\bibinfo{volume}{554}}, \bibinfo{pages}{55 } (\bibinfo{year}{2000}).

\bibitem[{\citenamefont{Yuan et~al.}(2015)\citenamefont{Yuan, Ju, Xie, Wang,
  and Chu}}]{Yuan2015164} 
\bibinfo{author}{\bibfnamefont{M.}~\bibnamefont{Yuan}},
  \bibinfo{author}{\bibfnamefont{X.}~\bibnamefont{Ju}},
  \bibinfo{author}{\bibfnamefont{R.}~\bibnamefont{Xie}},
  \bibinfo{author}{\bibfnamefont{W.}~\bibnamefont{Wang}}, \bibnamefont{and}
  \bibinfo{author}{\bibfnamefont{L.}~\bibnamefont{Chu}},
  \bibinfo{title}{Micromechanical properties of poly(N-isopropylacrylamide) hydrogel microspheres determined using a simple method},
  \bibinfo{journal}{Particuology} \textbf{\bibinfo{volume}{19}},
  \bibinfo{pages}{164 } (\bibinfo{year}{2015}).

\bibitem[{\citenamefont{Sierra-Mart\'{\i}n
  et~al.}(2011)\citenamefont{Sierra-Martin, Laporte, South, Lyon, and
  Fernandez-Nieves}}]{PhysRevE.84.011406}
\bibinfo{author}{\bibfnamefont{B.}~\bibnamefont{Sierra-Martin}},
  \bibinfo{author}{\bibfnamefont{Y.}~\bibnamefont{Laporte}},
  \bibinfo{author}{\bibfnamefont{A.~B.} \bibnamefont{South}},
  \bibinfo{author}{\bibfnamefont{L.~A.} \bibnamefont{Lyon}}, \bibnamefont{and}
  \bibinfo{author}{\bibfnamefont{A.}~\bibnamefont{Fern\'andez-Nieves}},
  \bibinfo{title}{Bulk modulus of poly($N$-isopropylacrylamide) microgels through the swelling transition},
  \bibinfo{journal}{Phys. Rev. E} \textbf{\bibinfo{volume}{84}},
  \bibinfo{pages}{011406} (\bibinfo{year}{2011}).

\bibitem[{\citenamefont{Sierra-Martin
  et~al.}(2011{\natexlab{a}})\citenamefont{Sierra-Martin, Frederick, Laporte,
  Markou, Lietor-Santos, and Fernandez-Nieves}}]{Sierra-Martin2011} 
\bibinfo{author}{\bibfnamefont{B.}~\bibnamefont{Sierra-Martin}},
  \bibinfo{author}{\bibfnamefont{J.~A.} \bibnamefont{Frederick}},
  \bibinfo{author}{\bibfnamefont{Y.}~\bibnamefont{Laporte}},
  \bibinfo{author}{\bibfnamefont{G.}~\bibnamefont{Markou}},
  \bibinfo{author}{\bibfnamefont{J.~J.} \bibnamefont{Lietor-Santos}},
  \bibnamefont{and}
  \bibinfo{author}{\bibfnamefont{A.}~\bibnamefont{Fernandez-Nieves}},
  \bibinfo{title}{Determination of the bulk modulus of microgel particles},
  \bibinfo{journal}{Colloid and Polymer Science}
  \textbf{\bibinfo{volume}{289}}, \bibinfo{pages}{721}
  (\bibinfo{year}{2011}{\natexlab{a}}).

\bibitem[{\citenamefont{Hashmi and Dufresne}(2009)}]{Hashmi:2009bv}
\bibinfo{author}{\bibfnamefont{S.~M.} \bibnamefont{Hashmi}} \bibnamefont{and}
  \bibinfo{author}{\bibfnamefont{E.~R.} \bibnamefont{Dufresne}},
  \bibinfo{title}{Mechanical properties of individual microgel particles through the deswelling transition},
  \bibinfo{journal}{Soft Matter} \textbf{\bibinfo{volume}{5}},
  \bibinfo{pages}{3682} (\bibinfo{year}{2009}).

\bibitem[{\citenamefont{Wyss et~al.}(2010)\citenamefont{Wyss, Franke, Mele, and
  Weitz}}]{Wyss:2010iu}
\bibinfo{author}{\bibfnamefont{H.~M.} \bibnamefont{Wyss}},
  \bibinfo{author}{\bibfnamefont{T.}~\bibnamefont{Franke}},
  \bibinfo{author}{\bibfnamefont{E.}~\bibnamefont{Mele}}, \bibnamefont{and}
  \bibinfo{author}{\bibfnamefont{D.~A.} \bibnamefont{Weitz}},
  \bibinfo{title}{Capillary micromechanics: Measuring the elasticity of microscopic soft objects},
  \bibinfo{journal}{Soft Matter} \textbf{\bibinfo{volume}{6}},
  \bibinfo{pages}{4550} (\bibinfo{year}{2010}).

\bibitem[{\citenamefont{Guo and Wyss}(2011)}]{Guo:2011di}
\bibinfo{author}{\bibfnamefont{M.}~\bibnamefont{Guo}} \bibnamefont{and}
  \bibinfo{author}{\bibfnamefont{H.~M.} \bibnamefont{Wyss}},
  \bibinfo{title}{Micromechanics of soft particles},
  \bibinfo{journal}{Macromol. Mater. Eng.} \textbf{\bibinfo{volume}{296}},
  \bibinfo{pages}{223} (\bibinfo{year}{2011}).
\bibitem[{\citenamefont{Voudouris et~al.}(2013)\citenamefont{Voudouris, Florea,
  van~der Schoot, and Wyss}}]{Voudouris:2013bc}
\bibinfo{author}{\bibfnamefont{P.}~\bibnamefont{Voudouris}},
  \bibinfo{author}{\bibfnamefont{D.}~\bibnamefont{Florea}},
  \bibinfo{author}{\bibfnamefont{P.}~\bibnamefont{van~der Schoot}},
  \bibnamefont{and} \bibinfo{author}{\bibfnamefont{H.~M.} \bibnamefont{Wyss}},
  \bibinfo{title}{Micromechanics of temperature sensitive microgels: Dip in the Poisson ratio near the LCST},
  \bibinfo{journal}{Soft Matter} \textbf{\bibinfo{volume}{9}},
  \bibinfo{pages}{7158} (\bibinfo{year}{2013}).
  
\bibitem[{\citenamefont{Hirotsu}(1990)}]{Hirotsu:1990uj}
\bibinfo{author}{\bibfnamefont{S.}~\bibnamefont{Hirotsu}},
\bibinfo{title}{Elastic anomaly near the critical point of volume phase transition in polymer gels},
\bibinfo{journal}{Macromolecules} \textbf{\bibinfo{volume}{23}},
\bibinfo{pages}{903} (\bibinfo{year}{1990}).

\bibitem[{\citenamefont{Gladman et al.}(2003)}]{SydneyGladman2016413}
\bibinfo{author}{\bibfnamefont{Sydney ~A.~Gladman}},
\bibinfo{author}{\bibfnamefont{E.~A.~Matsumoto}},
\bibinfo{author}{\bibfnamefont{R.~G.~Nuzzo}},
\bibinfo{author}{\bibfnamefont{L.~Mahadevan}},
\bibinfo{author}{\bibfnamefont{J.~A.~Lewis}},
\bibinfo{title}{Biomimetic 4D printing},
\bibinfo{journal}{Nature Materials} \textbf{\bibinfo{volume}{15}},
\bibinfo{pages}{413-418}(\bibinfo{year}{2003}).

\bibitem[{\citenamefont{Grinberg et~al.}(199)\citenamefont{Natalia V. Grinberg, Alexander S. Dubovik, Valerij Ya.
        Grinberg, Dmitry V. Kuznetsov, Elena E. Makhaeva, Alexander Yu. Grosberg, and Toyoichi Tanaka}}]{Grinberg1999}
\bibinfo{author}{\bibfnamefont{N. V.} \bibnamefont{Grinberg}},
\bibinfo{author}{\bibfnamefont{A. S.}~\bibnamefont{Dubovik}},
\bibinfo{author}{\bibfnamefont{V. Y.}~\bibnamefont{Grinberg}},
\bibinfo{author}{\bibfnamefont{D. V.} \bibnamefont{Kuznetsov}},
\bibinfo{author}{\bibfnamefont{E. E.}~\bibnamefont{Makhaeva}},
\bibinfo{author}{\bibfnamefont{A. Y.}~\bibnamefont{Grosberg}},
\bibnamefont{and}
\bibinfo{author}{\bibfnamefont{T.} \bibnamefont{Tanaka}},
\bibinfo{title}{Studies of the Thermal Volume Transition of
Poly(N-isopropylacrylamide) Hydrogels by High-Sensitivity Differential
Scanning Microcalorimetry. 1. Dynamic effects},
\bibinfo{journal}{Macromolecules}
\textbf{\bibinfo{volume}{32(5)}}, \bibinfo{pages}{1471-1475} (\bibinfo{year}{1999}).



\bibitem[{\citenamefont{Fernandez-Nieves et~al.}(2003)\citenamefont{Fernandez-Nieves, Fernandez-Barbero, Vincent, and
  de~Las~Nieves}}]{FernandezNieves:2003hh}
\bibinfo{author}{\bibfnamefont{A.}~\bibnamefont{Fernandez-Nieves}},
  \bibinfo{author}{\bibfnamefont{A.}~\bibnamefont{Fernandez-Barbero}},
  \bibinfo{author}{\bibfnamefont{B.}~\bibnamefont{Vincent}}, \bibnamefont{and}
  \bibinfo{author}{\bibfnamefont{F.~J.} \bibnamefont{de~Las~Nieves}},
 \bibinfo{title}{Osmotic de-swelling of ionic microgel particles},
 \bibinfo{journal}{J Chem Phys} \textbf{\bibinfo{volume}{119}},
  \bibinfo{pages}{10383} (\bibinfo{year}{2003}).

\bibitem[]{Supplemental}
\bibinfo{title}{See \textbf{Supplemental Material} at the end of this document for an overview of the additional DSC experiments, background information on the data analysis and the error analysis.}


\bibitem[{\citenamefont{Brian et~al.}(2003)\citenamefont{Brian R. Saunders and Brian Vincent}}]{Brain:1996}
\bibinfo{author}{\bibnamefont{ Brian R. Saunders}},
  \bibinfo{author}{\bibnamefont{Brian Vincent}}  ,
  \bibinfo{title}{Thermal and osmotic deswelling of poly(NIPAM) microgel particles},
  \bibinfo{journal}{J Chem SOC, Faraday Trans} \textbf{\bibinfo{volume}{92(18)}},
  \bibinfo{pages}{3385-3389} (\bibinfo{year}{1996}).

\bibitem[{\citenamefont{Bonnet-Gonnet et al}(1994)}]{Bonnet1944}
\bibinfo{author}{\bibfnamefont{C. Bonnet-Gonnet,}},
\bibinfo{author}{\bibfnamefont{L. Belloni,}},
\bibinfo{author}{\bibfnamefont{and}},
\bibinfo{author}{\bibfnamefont{B. Cabane,}},
\bibinfo{title}{Osmotic-Pressure Of Latex Dispersions},
\bibinfo{journal}{Langmuir} 
\textbf{\bibinfo{volume}{10}},
\bibinfo{pages}{4012-4021}(\bibinfo{year}{1994}).


\bibitem[{\citenamefont{Sierra-Martin
  et~al.}(2011{\natexlab{b}})\citenamefont{Sierra-Martin, Frederick, Laporte,
  Markou, Lietor-Santos, and Fernandez-Nieves}}]{SierraMartin:2011it}
\bibinfo{author}{\bibfnamefont{B.}~\bibnamefont{Sierra-Martin}},
  \bibinfo{author}{\bibfnamefont{J.~A.} \bibnamefont{Frederick}},
  \bibinfo{author}{\bibfnamefont{Y.}~\bibnamefont{Laporte}},
  \bibinfo{author}{\bibfnamefont{G.}~\bibnamefont{Markou}},
  \bibinfo{author}{\bibfnamefont{J.~J.} \bibnamefont{Lietor-Santos}},
  \bibnamefont{and}
  \bibinfo{author}{\bibfnamefont{A.}~\bibnamefont{Fernandez-Nieves}},
  \bibinfo{title}{Determination of the bulk modulus of microgel particles},
  \bibinfo{journal}{Colloid and Polymer Science}
  \textbf{\bibinfo{volume}{289}}, \bibinfo{pages}{721}
  (\bibinfo{year}{2011}{\natexlab{b}}).



\bibitem[{\citenamefont{Rubinstein and Colby}(2003)}]{rubinsteinbook}
\bibinfo{author}{\bibfnamefont{M.}~\bibnamefont{Rubinstein}} \bibnamefont{and}
  \bibinfo{author}{\bibfnamefont{R.H.} \bibnamefont{Colby}},
  \emph{\bibinfo{title}{Polymer physics}},
  \bibinfo{journal}{(Oxford University, New York)},
  (\bibinfo{year}{2003}).


\end{thebibliography}
\end{document}